\documentclass[final,3p,times,twocolumn]{elsarticle}
\usepackage{amsmath}
\usepackage{amssymb}
\usepackage{amsfonts}
\usepackage{amsthm}

\newtheorem{theorem}{Theorem}[section]
\newtheorem{corollary}[theorem]{Corollary}
\newtheorem{lemma}{Lemma}[section]
\newtheorem{definition}{Definition}[section]
\newtheorem{proposition}{Proposition}[section]
\newtheorem{example}{Example}[section]

\newcommand{\eqn}[1]{ \begin{equation} #1 \end{equation} }

\usepackage{graphicx}
\usepackage{color}

\newcommand{\transpose}{\intercal}

\journal{Physics Letters B}

\newcommand{\preprint}{
 \setlength{\unitlength}{1mm}{\hbox{\begin{picture}(0,0)
      \put(160,10){\mbox{\footnotesize%
        ADP-12-38/T805}}\end{picture}}}}

\begin{document}

\begin{frontmatter}

\title{\preprint Wave Function of the Roper from Lattice QCD}

\author[adl]{Dale S.~Roberts}
\author[adl]{Waseem Kamleh}
\author[adl]{Derek B.~Leinweber}
\address[adl]{Special Research Centre for the Subatomic Structure of
  Matter, School of Chemistry and Physics, University of Adelaide, SA,
  5005, Australia} 

\begin{abstract}
We apply the eigenvectors from a variational analysis in lattice QCD
to successfully extract the wave function of the Roper state, and
  a higher mass $P_{11}$ state of the nucleon.  We use the $2+1$
flavour $32^3\times 64$ PACS-CS configurations at a near physical pion
mass of 156 MeV.  We find that both states exhibit a structure
consistent with a constituent quark model. The Roper $d$-quark wave
function contains a single node consistent with a $2S$ state, and
the third state wave function contains two, consistent with a
$3S$ state. A detailed comparison with constituent quark model wave
functions is carried out, obtained from a Coulomb plus ramp
potential. These results validate the approach of accessing these
states by constructing a variational basis composed of different
levels of fermion source and sink smearing.  Furthermore, significant
finite volume effects are apparent for these excited states which mix
with multi-particle states, driving their masses away from physical
values and enabling the extraction of resonance parameters from
lattice QCD simulations.
\end{abstract}

\end{frontmatter}

\section{Introduction}

The wave function and associated probability distribution of a
particle in a potential are fundamental to the very nature of quantum
mechanics.  In the non-relativistic case, the entire spectrum of the
particle can be determined by solving the Schr\"{o}dinger equation. In
quantum field theory, a Schr\"{o}dinger-like probability distribution
can be constructed for bound states by analogy, by taking a
  simplified view of the full quantum field theory wave functional in
the form of the Bethe-Salpeter wave function \cite{Salpeter:1951sz},
herein referred to as simply the 'wave function'.

Recent advances in the isolation of nucleon excited states through
correlation-matrix based variational techniques in lattice QCD now
enable the exploration of the structure of these states and how these
properties emerge from the fundamentals of QCD.  In this letter, we
report the first results for the wave function of the first
even-parity excitation of the nucleon, the Roper \cite{Roper:1964zza}.  

Early explorations of these states considered a non-relativistic
constituent quark model.  The probability distributions of quarks
within hadrons were determined using a one-gluon-exchange potential
augmented with a confining form \cite{DeRujula:1975ge,Bhaduri:1980fd}.
These models have been the cornerstone of intuition of hadronic
probability distributions for many decades, and have been complemented
with features such as meson-cloud dressing.

In this investigation, we will confront these early predictions for
quark probability distributions in excited states directly via Lattice
QCD.  Visualizations of the probability distributions for ground
states on the lattice \cite{Velikson:1984qw} have been used to observe
interesting physical effects such as Lorentz contraction
\cite{Chu:1990ps,Gupta:1993vp}, quarks aligning with a magnetic field
and diquark clustering \cite{Roberts:2010cz}. Furthermore, the
probability distribution can be used as a diagnostic tool, allowing
finite volume effects and other lattice artifacts to be easily
visualized and understood \cite{Alexandrou:2008ru}.

The Bethe-Salpeter wave function underlying the probability
distributions can be defined in the form of a gauge-invariant
Bethe-Salpeter amplitude.  For the wave function of the $d$ quark
about two $u$ quarks in the proton, $\vert p \rangle$, the amplitude
takes the form
\begin{eqnarray}
\psi_d^p(y) &\propto& \int d^4x\, \langle \Omega \vert \, \epsilon^{abc} \, 
u^{\transpose a}(x)\, C \gamma_5\,  \nonumber \\
&&\left [P \exp{\left ( ig \int_x^{x+y} A(x^\prime) \cdot
  dx^\prime \right )}\,  d(x+y) \right ]^b \nonumber \\
&& u^c(x)
\, \vert p \rangle \, ,
\label{gaugeInvBSamp}
\end{eqnarray}
which exploits a string of flux to connect the quarks in a gauge
invariant manner.  Here we have selected the standard form of the
proton interpolating field $\chi_1$, $u$ and $d$ represent the up and
down quark fields respectively with colour indices $a$, $b$ and $c$
and $C$ is the charge conjugation matrix.

In a relativistic gauge theory the concept of a hadronic wave function
is not unique.  For example, in the gauge invariant form there is an
explicit path dependence.  For large separations of the quarks an
average over the paths is desirable.  This leads us to consider other
Bethe-Salpeter amplitudes in which the gauge degree of freedom is
fixed to a specific gauge.  In lattice field theory, Coulomb and
Landau gauges are most common due to their local gauge fixing
procedure.

Landau gauge is a smooth gauge that preserves the Lorentz invariance
of the theory.  It is a popular choice in the field and we select it
here.  While the size and shape of the wave function are gauge
dependent, our selection of Landau gauge is vindicated in
Sec.~\ref{sec:SimRes}.  There we illustrate how the ground state wave
function of the $d$ quark in the proton can be described accurately by
the non-relativistic quark model using standard values for the
constituent quark masses and string tension of the confining
potential.  Thus, Landau gauge provides an excellent forum for the
examination of the wave functions of the excited states of the proton
and their detailed comparison with traditional non-relativistic quark
model predictions.

\section{Lattice Techniques}

Hadron spectroscopy is a highly complex problem. Though it is
relatively simple to see higher energy resonances of hadrons in
colliders, apart from simple quantum numbers, properties more
fundamental to the nature of these resonances remain elusive to
experiment.

Robust methods have been developed that allow the isolation and study
of states associated with these resonances in Lattice QCD
\cite{Leinweber:1994nm,Gockeler:2001db,Sasaki:2001nf,Melnitchouk:2002eg,Lee:2002gn,Leinweber:2004it,Basak:2007kj,Bulava:2010yg,Mahbub:2010rm,Mahbub:2011zz}.
In this study, we apply the variational method
\cite{Michael:1985ne,Luscher:1990ck} to extract the ground state and
first two $P_{11}$ excited states of the proton associated with the
Roper \cite{Roper:1964zza} and another $P_{11}$ state.  We then
combine this with lattice wave function techniques to calculate the
probability distributions of these states at near-physical quark
masses.  We use the $2+1$ flavour $32^3\times 64$ PACS-CS
configurations \cite{Aoki:2008sm} at a pion mass of 156 MeV.

The wave function of a hadron is proportional to the parity-projected
\cite{Lee:1998cx} two-point Green's
function, 
\eqn{G_{ij}^\pm(\vec{p},t) = \sum_{\vec x}
  e^{-i\vec{p}\cdot\vec{x}} 
\, \mbox{tr}\, \left ( \gamma_0 \pm 1 \right ) \,
\langle\Omega\vert \,
  T\{\, \chi_i(\vec{x},t)\, \bar\chi_j(0,0)\, \}\, \vert\Omega\rangle
  \, , 
\label{twopt}}
where $\chi_i$ are the hadronic interpolating fields. In the case of
the proton the most commonly used interpolator is given by 
\eqn{\chi_1(x)=\epsilon^{abc}\,
  (\, u^{\transpose a}(x)\, C\gamma_5\, d^b(x)\, )\, u^c(x) \, ,} 
with the corresponding adjoint given by
\eqn{\bar\chi_1(0)=\epsilon^{abc}\,
  (\, \bar{d}^b(0)\, C\gamma_5\, \bar{u}^{\transpose a}(0)\, )\,
  \bar{u}^c(0) \, .}
In order to construct the wave function, the quark fields in the
annihilation operator are each given a spatial dependence, 
\eqn{\chi_1(\vec{x},\vec{y},\vec{z},\vec{w}) = \epsilon^{abc}\, (\,
  u^{\transpose a}(\vec{x}+\vec{y})\, C\gamma_5\, d^b(\vec{x}+\vec{z})\, )\,
  u^c(\vec{x}+\vec{w}) \, , \label{gdepinterp} } 
while the creation operator remains local.  This generalizes $G(\vec
p, t)$ to a wave function proportional to $G(\vec p, t; \vec y, \vec
z, \vec w)$. In principle, we could allow each of these coordinates,
$\vec y,\ \vec z,\ \vec w$, to vary across the entire lattice,
however, we can reduce the complexity by taking advantage of the
hyper-cubic rotational and translational symmetries of the lattice and
considering the system centre of mass.  A description of the
probability distribution of a particular quark within the proton can
be formed by holding the spatial location of two of the quarks fixed
and calculating the third quark's amplitude at every lattice site. We
focus on the probability distribution of the $d$ quark from
Eq.~(\ref{gdepinterp}), with the $u$-quarks fixed at the origin, i.e.
\eqn{\chi_1(\vec{x},0,\vec{z},0;t) = \epsilon^{abc}\, (\,
  u^{\transpose a}(\vec{x},t)\, C\gamma_5\, d^b(\vec{x}+\vec{z},t)\, )\,
  u^c(\vec{x},t) \, . \label{dqinterp} } 
The coordinate $\vec x$ then represents the centre of the system, providing
an origin analogous to a potential well in the non-relativistic case.

The variational method \cite{Michael:1985ne,Luscher:1990ck} is a
well-established method \cite{Mahbub:2009nr} for extracting the
excited state spectra of hadrons.  Noting that the only time
dependence in the two-point function lies in the exponential, we are
able to construct the following relation
\eqn{G_{ij}(t_0+\Delta t)\, u_j^\alpha = e^{-m_\alpha\Delta t}\,
  G_{ij}(t_0)\, u^\alpha_j
  \, , 
\label{eqn:rightEVeq} }
where $u^\alpha$ are the right eigenvectors of the eigenvalue equation,
\eqn{(\,G^{-1}(t_0)\, G(t_0+\Delta t)\, )_{ij}\,
  u^\alpha_j=e^{-m_\alpha\Delta t} \, u^\alpha_i.}

Similarly, we can construct the left eigenvector equation
\eqn{
v_i^\alpha\, (\, G(t_0+\Delta t)\, G^{-1}(t_0)\, )_{ij} =
e^{-m_\alpha\Delta t} \, v_j^\alpha \, .
}
To project a single state, one applies the eigenvectors to the
parity-projected variational matrix
\eqn{v_i^\alpha\, G_{ij}^+(t)\, u^\beta_j\propto \delta_{\alpha\beta}\,
  e^{-m_\alpha^+ \, \Delta t} \, .
\label{standard_tech}} 
The effective mass can then be calculated from the projected two-point
functions as $m(t) = \log \left ( G(t) / G(t+1) \right )$.  While the
effective mass is insensitive to a wide range of variational
parameters \cite{Mahbub:2010rm}, we follow Ref.~\cite{Mahbub:2010rm}
and select $t_0$ to be 2 time slices after the source with $\Delta
t=2$.

Different interpolators exhibit different couplings to the proton
ground and excited states and hence can be used to construct a
variational basis. The limited number of local interpolators restricts
the size of the operator basis \cite{Leinweber:1994nm}. To remedy
this, one can exploit the smearing dependence of the coupling of
states to one or more standard interpolating operators in order to
construct a larger variational basis where the $\chi_i$ and
$\bar\chi_j$ from Eq.~(\ref{twopt}) contain a smearing
dependence. This method has been shown to allow access to states
associated with resonances such as the Roper, $N^\star(1710)$
\cite{Mahbub:2010rm} and the $\Lambda(1405)$ \cite{Menadue:2011pd}.

\begin{figure}[t]
\includegraphics[angle=90,width=0.98\linewidth]{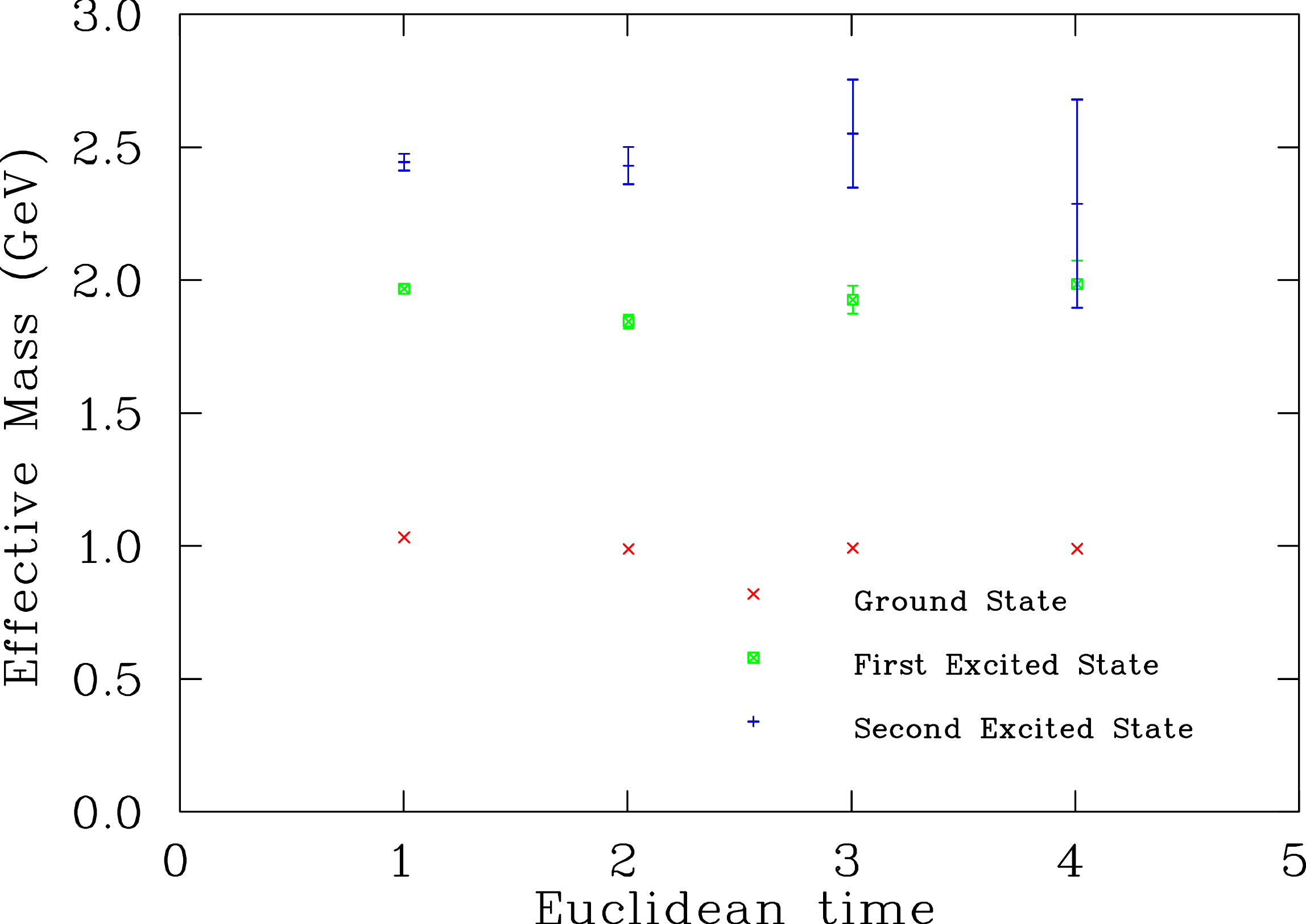}
\caption{The effective mass of the ground, Roper, and a higher
    excited state of the nucleon projected with only the right
  eigenvector. Effective projection is accomplished up to 3 time
  slices after the source. }
\label{ProjEffMass}
\end{figure}

The non-local sink operator used to construct the wave function is
unable to be smeared, such that the standard technique of
Eq.~(\ref{standard_tech}) cannot be applied.  However,
Eq.~(\ref{eqn:rightEVeq}) illustrates it is sufficient to isolate the
state at the source using the right eigenvector.  Thus, the
probability distributions are calculated with each smeared source
operator and the right eigenvectors calculated from the standard
variational analysis are then applied in order to extract the
individual states.  As demonstrated in Fig.~\ref{ProjEffMass}, clean
projection of two excited states is obtained.  We note how the
plateaus commence at $t=t_0=2$, where the correlation matrix analysis
has been applied.  As the fourth state may accommodate a superposition
of all remaining spectral strength in the correlator, we do not
consider it further.

Our focus on $\chi_1$ in this investigation follows from the results
of Ref.~\cite{Mahbub:2013ala}, where the lowest-lying excitation of
the nucleon was shown to be predominantly associated with the $\chi_1$
interpolating field.  The results from their $8 \times 8$ correlation
matrix of $\chi_1$ and $\chi_2 = \epsilon^{abc}\, (\, u^{\transpose
  a}(x)\, C\, d^b(x)\, )\, \gamma_5 \, u^c(x)$ revealed that $\chi_2$
plays a marginal role in exciting the Roper.  The coefficients
of the Roper source eigenvector multiplying $\overline\chi_2$ are near
zero.  Instead this interpolating field is key to obtaining good
overlap with the $N(1710)$ excited state of the nucleon.  Further
comparison with Ref.~\cite{Mahbub:2013ala}, identifies the third state
extracted herein as the fifth state of the $8\times 8$ analysis,
illustrated in Fig.~6 of Ref.~\cite{Mahbub:2013ala} as the green star
at the lightest quark mass.

In summary, the wave function for the $d$ quark in state $\alpha$
having momentum $\vec p$ observed at Euclidean time $t$ is
\begin{align}
\psi^\alpha_d&(\vec p, t; \vec z) =
\sum_{\vec x}
  e^{-i\vec{p}\cdot\vec{x}}  \label{psi_alpha} \\
&\mbox{tr}\, \left ( \gamma_0 \pm 1 \right ) \,  \langle\Omega\vert
  \,T\{\, \chi_1(\vec{x},0,\vec z,0;t)\, \bar\chi_j(0,0)\,  \}\, 
  \vert\Omega\rangle \,
  u_j^\alpha \, , \nonumber
\end{align}
where $\chi_1(\vec{x},0,\vec z,0;t)$ is given by Eq.~(\ref{dqinterp}). 

As discussed above, $\chi_1$ has the spin-flavour construct that is
most relevant to the excitation of the Roper from the QCD vacuum.  As
such, it is an ideal choice for revealing the spatial distribution of
quarks within the Roper.  However, the selection of $\chi_1$ in 
Eq.~(\ref{psi_alpha}) is not unique and other choices are
possible.  For example, the selection of $\chi_2$ would reveal small
contributions to the Roper wave function where vector diquark degrees
of freedom are manifest.  Similarly, $D$-wave contributions could be
resolved through the consideration of a spin-3/2 isospin-1/2
interpolating field at the sink.

In carrying out our calculations, we average over the equally
weighted $\{ U \}$ and $\{ U^* \}$ link configurations as an improved
unbiased estimator.  The two-point function is then perfectly
real and the probability density is proportional to the square of the
wave function.  In this analysis, we choose to look at the
zero-momentum probability distributions three time slices after the
source.

\section{Simulation Results}
\label{sec:SimRes}

We use the $2+1$ flavour $32^3\times 64$ PACS-CS configurations
\cite{Aoki:2008sm}, constructed with the Iwasaki gauge action
\cite{Iwasaki:2011jk} with $\beta=1.90$, giving a lattice spacing of
$0.0907(13)\,\mathrm{fm}$, and the $\mathcal{O}(a)$-improved Wilson
action \cite{Sheikholeslami:1985ij}.  We use $198$ gauge field
configurations, and employ multiple sources per configuration,
separated by at least one quarter of the temporal lattice extent. The
hopping parameter for the light quarks is $\kappa_{ud}=0.13781$,
giving a pion mass of $156\,\mathrm{MeV}$.

To cleanly access the first three states, a $4\times 4$ variational
basis is constructed using the $\chi_1$ operator with $16$, $35$,
$100$ and $200$ sweeps of Gaussian smearing \cite{Gusken:1989qx},
corresponding to RMS smearing radii of $2.37$, $3.50$, $5.92$ and
$8.55$ lattice units respectively. We fix to Landau gauge by
maximizing the $\mathcal{O}(a^2)$ improved fixing functional
\cite{Bonnet:1999mj}
\begin{equation}
\mathcal{F}_{Imp} =\sum_{x,\mu}
\mathrm{Re}\,\mathrm{tr}\left(\frac{4}{3} U_\mu(x) -
\frac{1}{12\,u_0}\left(U_\mu(x)U(x+\hat{\mu}) + \mathrm{h.c.}\right)\right) 
\end{equation}
using a Fourier transform accelerated algorithm \cite{Davies:1987vs}.

The wave functions observed for all our states show an approximate
symmetry over the eight octants surrounding the origin.  To improve
our statistics we average over these eight octants before presenting
the results.

Our point of comparison with previous models of quark probability
distributions comes from a non-relativistic constituent quark model
with a one-gluon-exchange motivated Coulomb\,+\,ramp potential.  The
spin dependence of the model is given in Ref.~\cite{Bhaduri:1980fd}
and the radial Schrodinger equation is solved with boundary conditions
relevant to the lattice data; {\it i.e.} the derivative of the
wave function is set to vanish at a distance $L_x/2$.

\begin{figure}[t]
\includegraphics[width=0.98\linewidth]{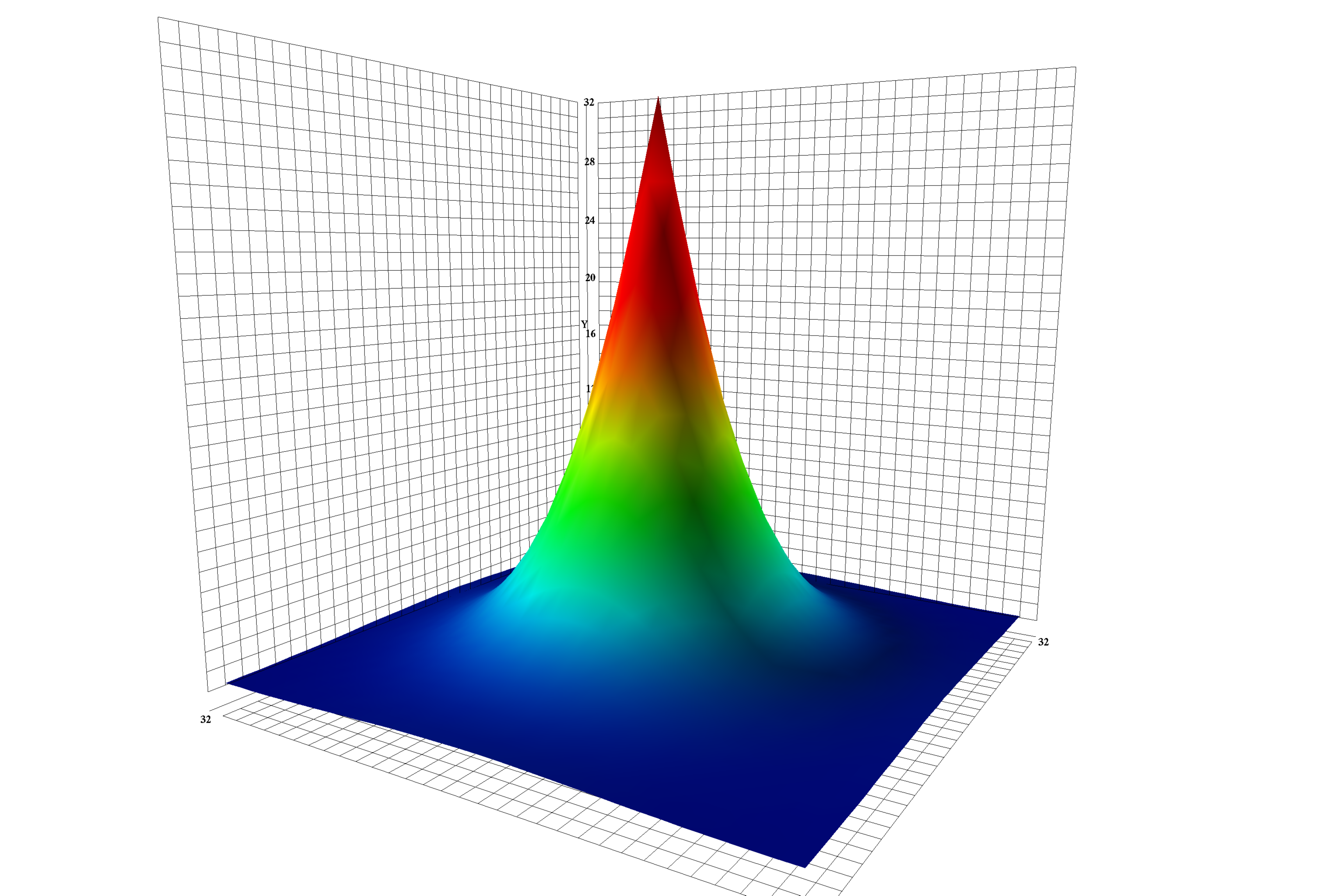}
\caption{The probability distribution of the ground state $d$ quark
  about the two $u$ quarks fixed at the origin. }
\label{SurfPlotS1}

\end{figure}

The ground state probability density for the $d$ quark about the two
$u$ quarks at the origin obtained in our lattice calculations is
illustrated in Fig.~\ref{SurfPlotS1}.  We see that the well-known
sharp-peaked shape associated with the Coulomb potential is
reproduced.

\begin{figure}[b!]
\includegraphics[angle=90,width=0.98\linewidth]{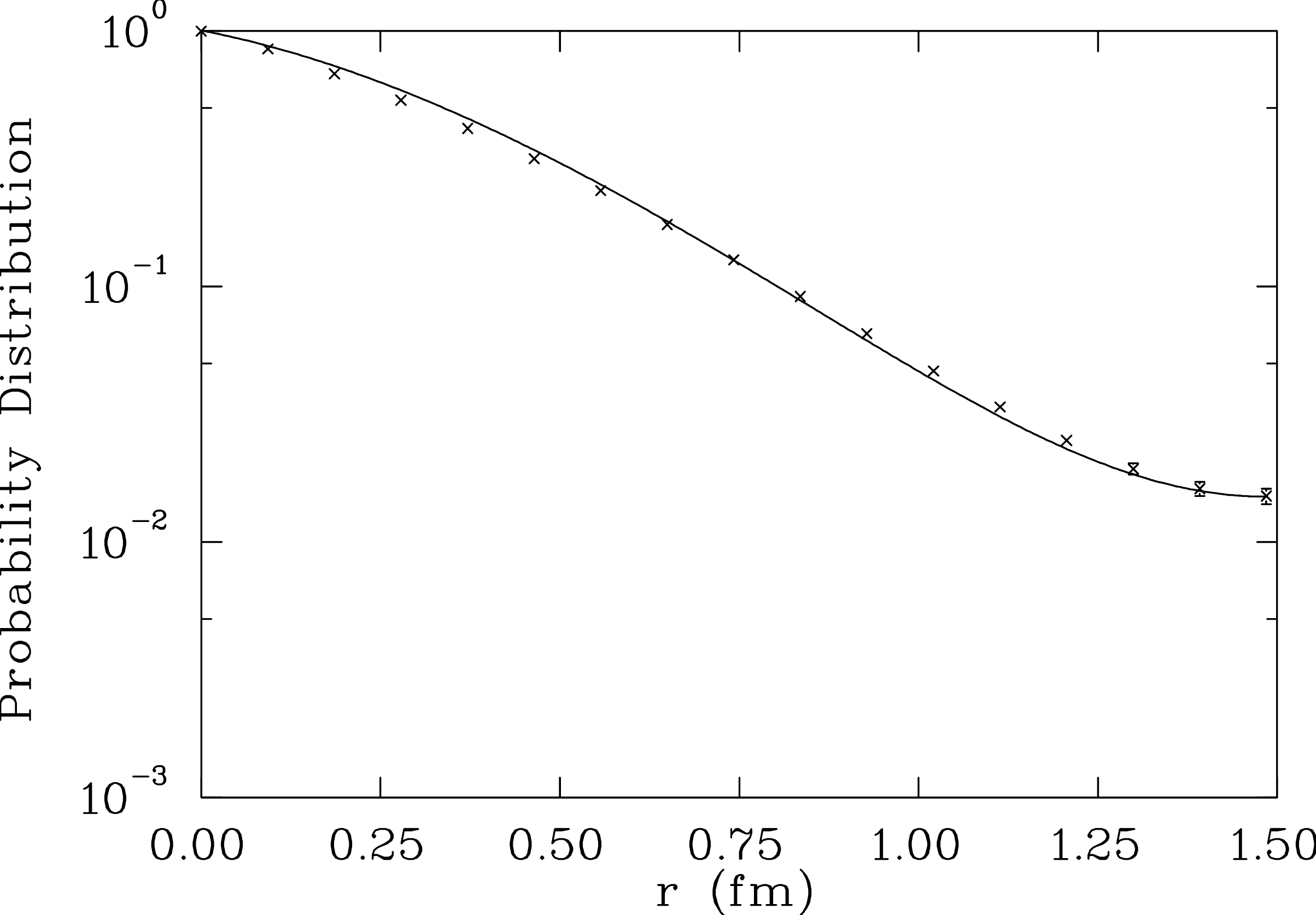}
\caption{Comparison of the ground state lattice probability
  distribution with the quark model.  The probability distributions
  are qualitatively similar, as would be expected for the ground
  state. }
\label{QMcompS1}
\end{figure}

The lattice data are compared with the constituent quark model in
Fig.~\ref{QMcompS1}.  Here, both the quark model probability
distributions and the lattice results have been scaled such that the
peak value is 1.  The two quark model parameters adjusted in the fit
are the string tension, $\sqrt{\sigma} = 440\pm 40$ MeV, and the
constituent quark mass, $m_q \sim 370$ MeV (accommodating the fact our
quark mass is above the physical value). Using a least-squares fit
varying the parameters $m_q$ and $\sqrt{\sigma}$, we find the ground
state lattice results are described well with $\sqrt{\sigma} = 400$
MeV and $m_q = 360$ MeV, which gives a ground state mass of
$940\,\mathrm{MeV}$ and a first excited state mass of
$1573\,\mathrm{MeV}$.  These parameters are held fixed in examinations
of the excited states.

\begin{figure}[t!]
\includegraphics[width=0.98\linewidth]{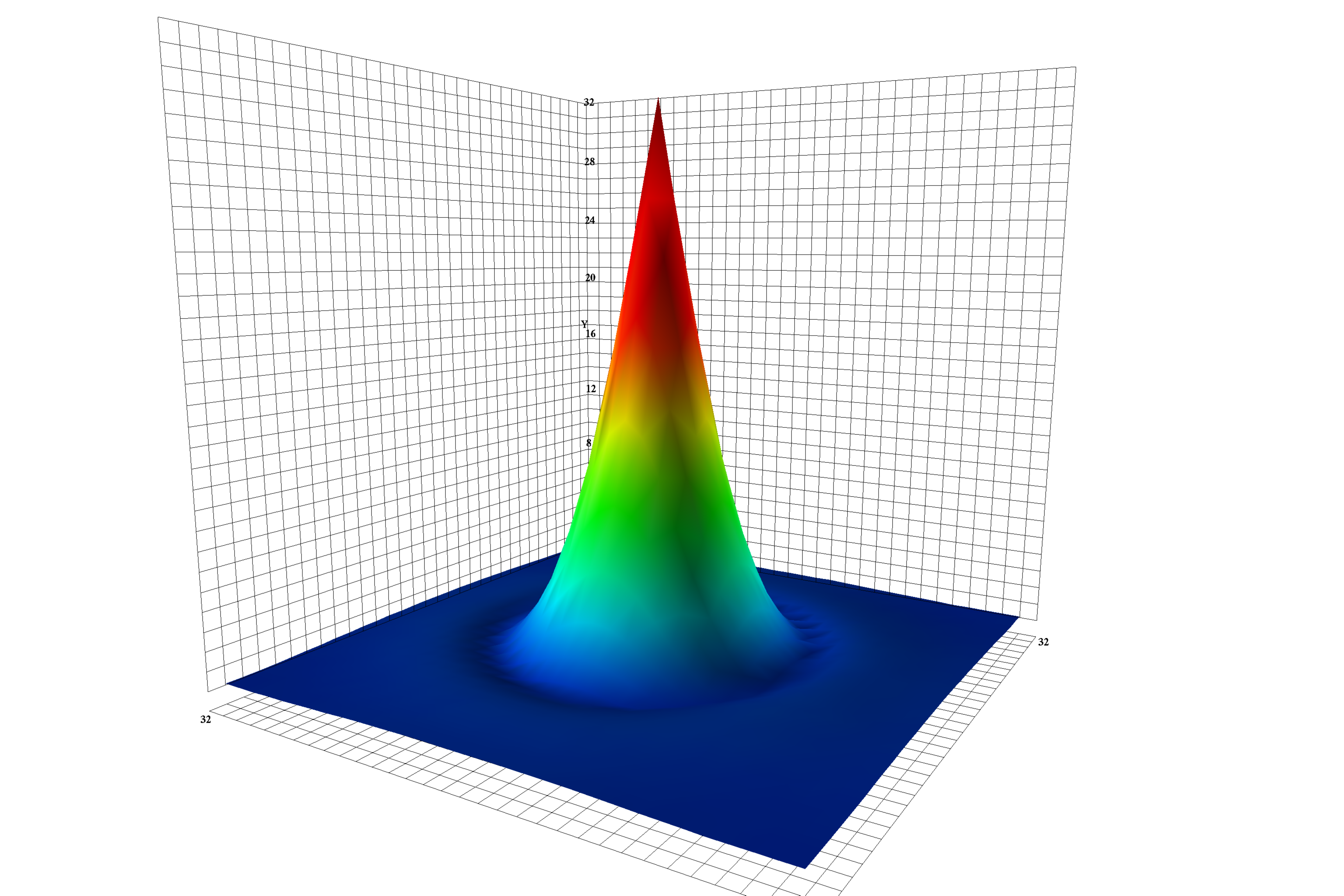}
\caption{The probability distribution of the $d$ quark about the two
  $u$ quarks at the origin in the first excited state. The darkened
  ring around the peak indicates a node in the probability
  distribution, consistent with a $2S$ state.}
\label{SurfPlotS2}
\end{figure}

Lattice results for the $d$-quark probability distribution in the
first excited state of the proton are
presented in Fig.~\ref{SurfPlotS2}.  The distribution exhibits a hydrogenic node structure
consistent with a $2S$ state, indicating that the state includes a radial
excitation of the $d$ quark. This structure also indicates that the ideal
combination of operators to access this state on the lattice would be
superposed Gaussians of different widths and opposite signs. This
observation validates the approach of combining multiple smearing
levels to construct the variational basis and indeed the alternating
signs of superposed Gaussians are observed in
Refs.~\cite{Mahbub:2011zz,Mahbub:2013ala}.

\begin{figure}[t!]
\includegraphics[width=0.98\linewidth]{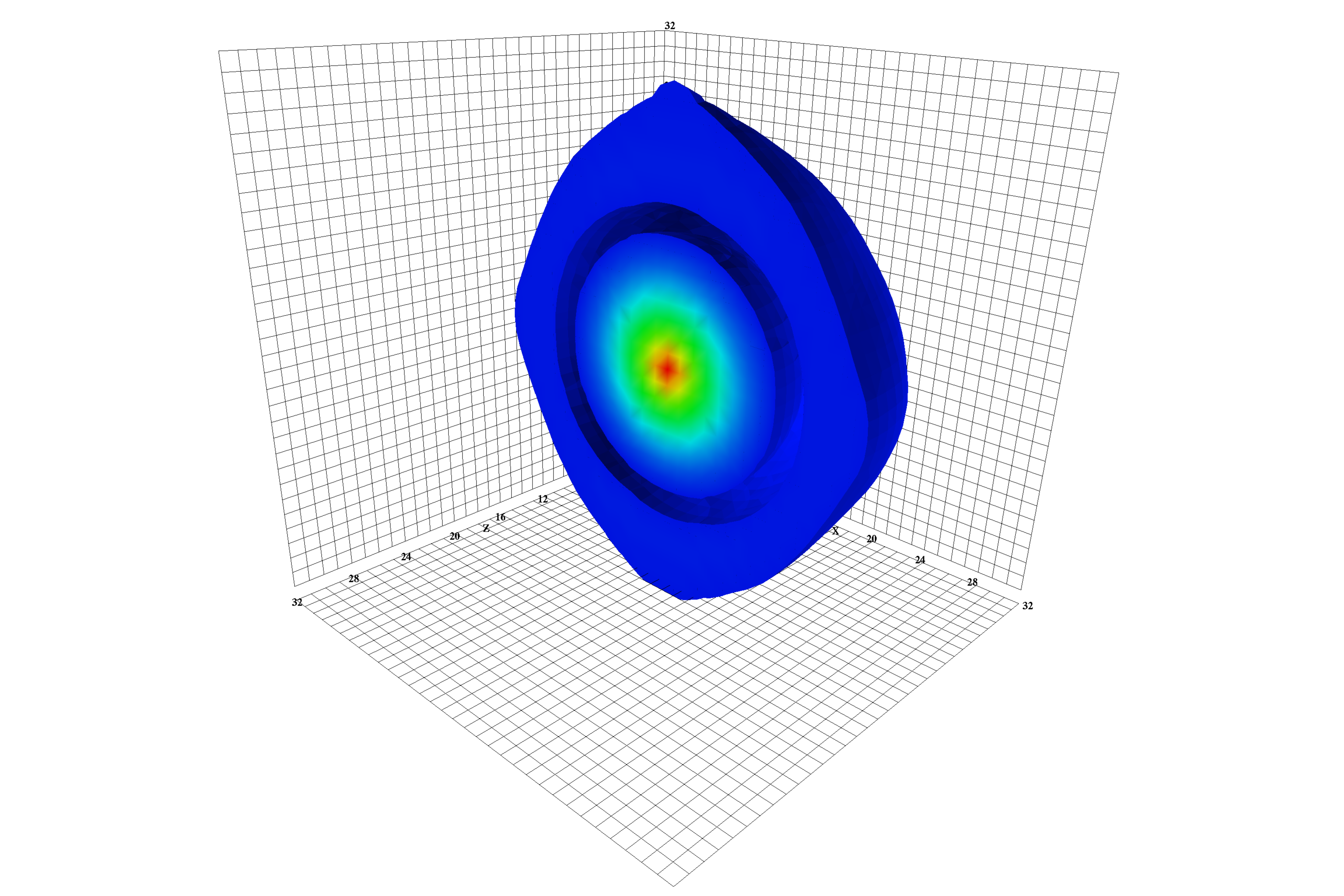}
\caption{The isovolume of the probability distribution of the $d$
  quark in the first excited state (colour map as in
  Fig.~\ref{SurfPlotS2}). The outer edge can be seen to be affected
  by the boundary, indicating a necessary finite-volume effect
  associated with multi-particle components of the state. }
\label{IsovolS2}
\end{figure}

The isovolume of this probability distribution illustrated in
Fig.~\ref{IsovolS2} clearly shows the nodal structure, with an inner
sphere surrounded by a near-spherical shell. The deviation from
spherical symmetry in the outer shell directly displays the important
interplay between the energy of the excited state observed in the
lattice simulation and the finite volume of the lattice.  At this very
light quark mass the distortion of the probability density is
significant and will correspondingly influence the lattice hadron
mass.  This interplay is key to extracting resonance parameters from
lattice simulation results.

\begin{figure}[b!]
\includegraphics[angle=90,width=0.98\linewidth]{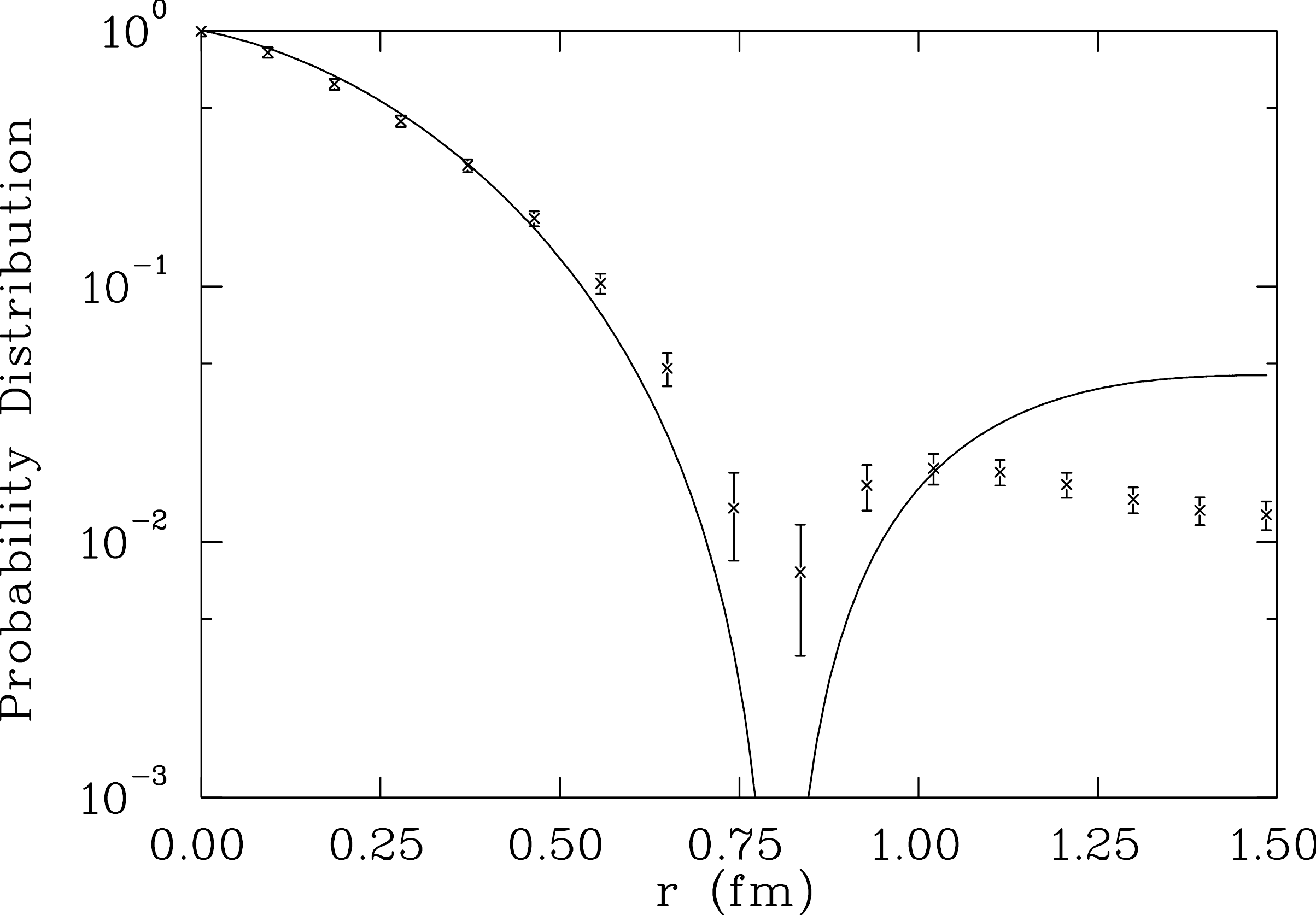}
\caption{Comparison of the first excited state $d$-quark probability
  distribution from our lattice QCD calculation (crosses) with the
  quark model (solid curve). The quark model predicts the node in
  approximately the correct location, but deviates at the boundary.}
\label{QMcompS2}
\end{figure}

Comparing the lattice probability distribution for the $d$ quark
in the first excited state to that predicted by the constituent quark model in
Fig.~\ref{QMcompS2}, we see a qualitative similarity but with
important differences. While the node position is similar, 
the shape of the wave function tail is different.

\begin{figure}[t!]
\includegraphics[width=0.98\linewidth]{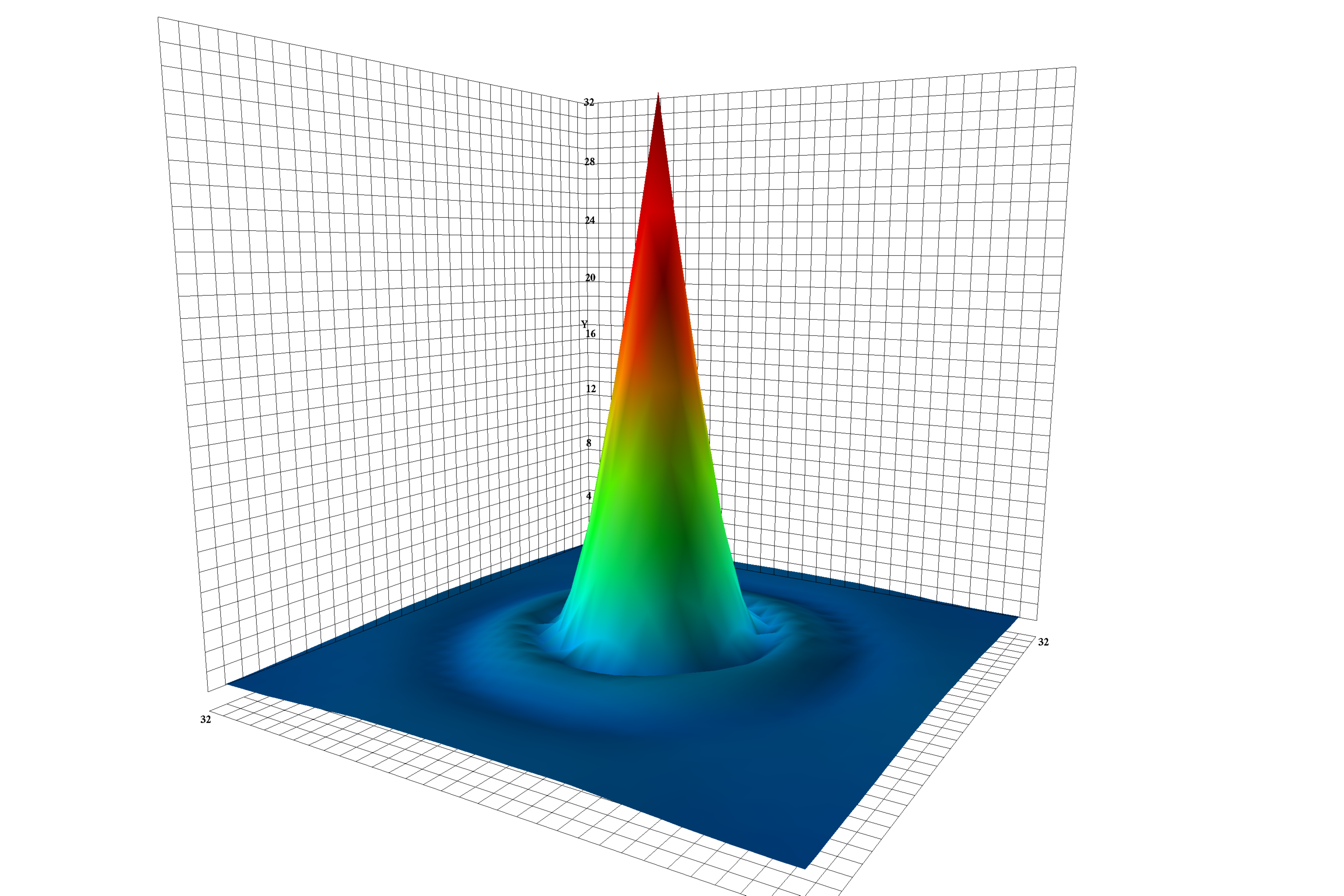}
\caption{The probability distribution of the $d$ quark in the second
  excited state of the nucleon. Two nodes are visible, consistent with
  a $3S$ state. }
\label{SurfPlotS3}
\end{figure}

\begin{figure}[t!]
\includegraphics[width=0.98\linewidth]{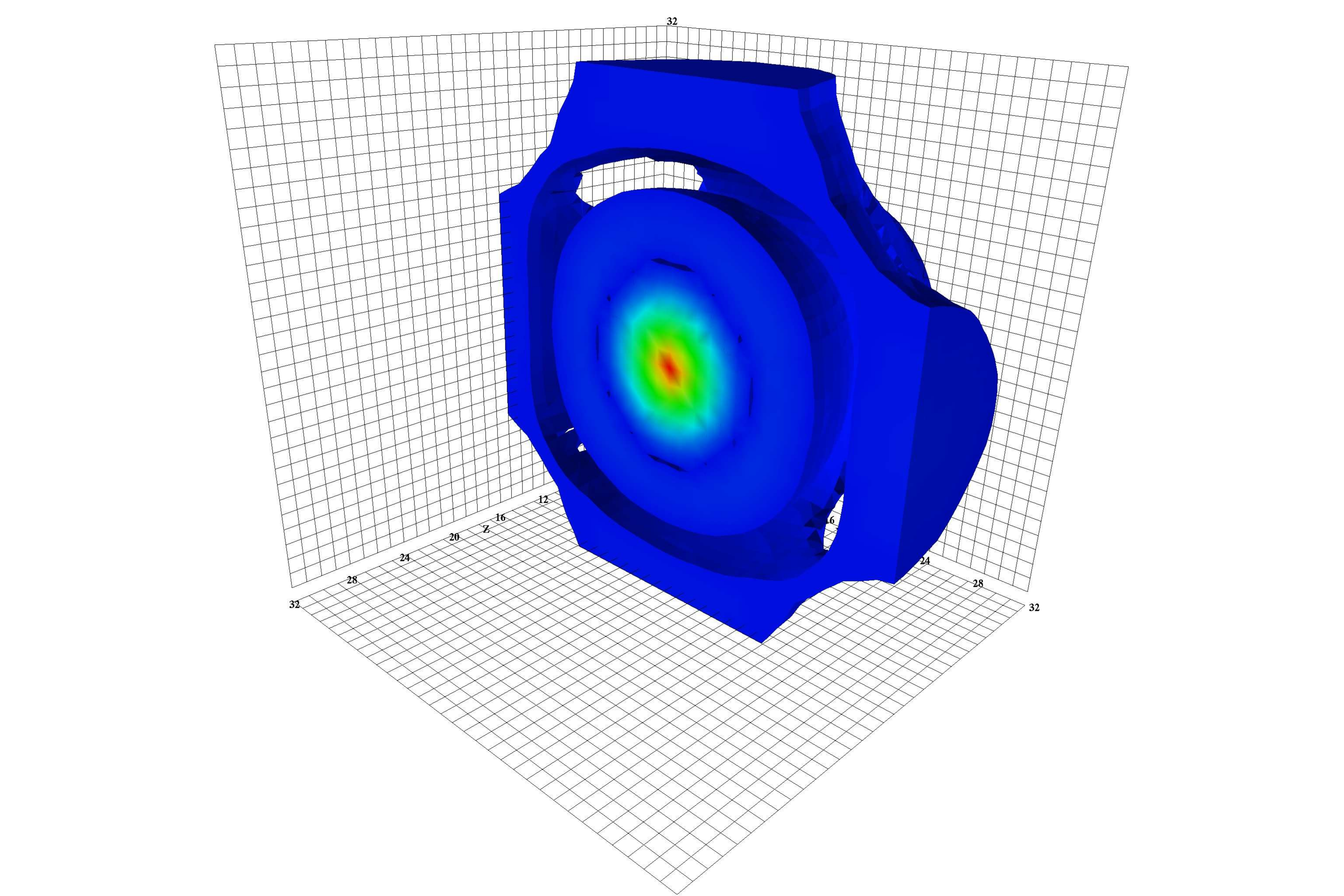}
\caption{The isovolume of the probability distribution of the $d$
  quark in the second excited state. The outermost node is compressed
  by the boundary into an almost square shape, indicating strong
  finite-volume effects.}
\label{IsovolS3}
\end{figure}

\begin{figure}[t!]
\includegraphics[angle=90,width=0.98\linewidth]{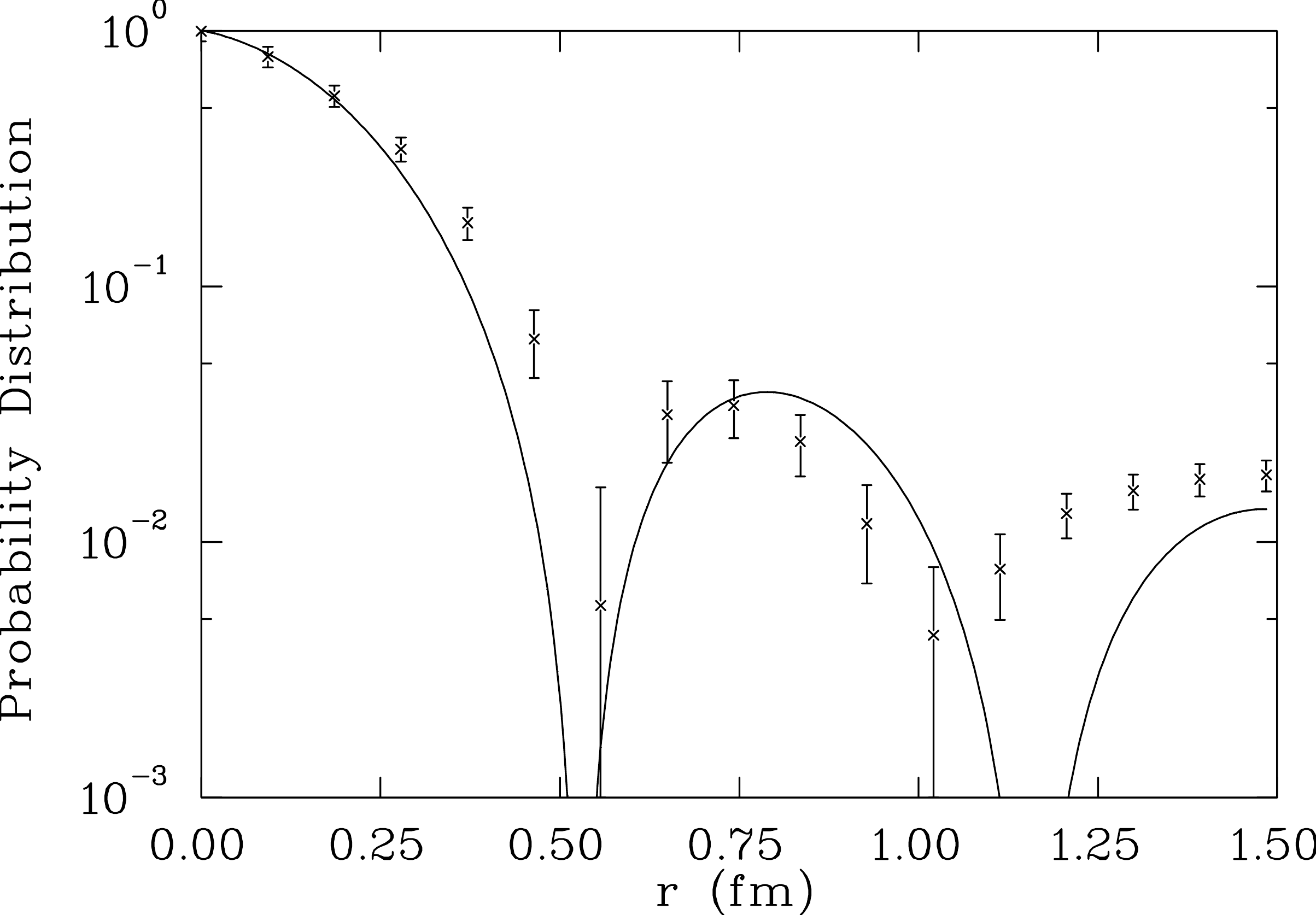}
\caption{Comparison of the second excited state $d$-quark probability
  distribution from the lattice (crosses) with the quark model (solid
  curve). The nodes in the lattice data fall in between those predicted
by the quark model.}
\label{QMcompS3}
\end{figure}

The probability distribution of the second excited state of the
nucleon in Fig.~\ref{SurfPlotS3} reveals two nodes consistent
with a $3S$ radial excitation of the $d$ quark.  Finite volume effects
become even more apparent as shown in Fig.~\ref{IsovolS3}, distorting the
outermost shell of the wave function into an almost square
shape. Comparing this state to the quark model prediction in
Fig.~\ref{QMcompS3}, we observe qualitative agreement.

\section{Summary}

In this world-first study of the quark probability distribution within
excited states of the nucleon, we have shown that both the Roper and
the second excited state examined herein display the node structure
associated with radial excitations of the quarks.  On comparing these
probability distributions to those predicted by a constituent quark
model, we find good qualitative similarity with interesting
differences.  The discovery of a node structure provides a deep
understanding of the success of the smeared-source/sink correlation
matrix methods of Ref.~\cite{Mahbub:2010rm}.

Finite volume effects were shown to be particularly significant for
the excited states explored herein at relatively light quark mass.  As
these excited states have a multi-particle component, the interplay
between the lattice volume, the wave function and the associated
energy are key to extracting the resonance parameters of the Roper.

Future calculations will explore the structure of the Roper in more
detail, examining the mass dependence of the wave functions, more
general spatial configurations of the quark positions, and the
introduction of isospin-1/2 spin-3/2 interpolating fields to reveal
the role of $D$-wave contributions to the Roper.  While our use of
improved actions suppresses lattice discretisation errors, ultimately
simulations will be done at a variety of lattice spacings directly at
the physical quark masses to connect the lattice QCD simulations to
the continuum results of Nature.

\section*{Acknowledgements}
This research was undertaken on the NCI National Facility in Canberra,
Australia, which is supported by the Australian Commonwealth
Government.  We also acknowledge eResearch SA for support of our
local supercomputing resources.  This research is supported by the
Australian Research Council.

\end{document}